\documentclass[aps,prl,reprint,showpacs,floatfix]{revtex4-2}

\usepackage{multirow}
\usepackage{mathtools}
\usepackage{amssymb}
\usepackage{amsthm}
\usepackage{eurosym}
\usepackage{textcomp}	%For Yen-symbol
\usepackage{dsfont}		%Double Stroke math font 
\usepackage{bm}
\usepackage{hyperref}
\usepackage{tikz}
		\usetikzlibrary{matrix}
\usepackage{soul}	%To highlight text, to be removed in final version

\theoremstyle{definition}

\newcommand{\ncom}{\newcommand}
\ncom{\h}{\mathcal{H}}
\ncom{\ket}[1]{\left|#1\right\rangle}
\ncom{\bra}[1]{\left\langle#1\right|}
\ncom{\braket}[2]{\left\langle#1\middle|#2\right\rangle}
\ncom{\ketbra}[1]{\left|#1\middle\rangle\middle\langle#1\right|}
\ncom{\expv}[2]{\left\langle#1\middle|#2\middle|#1\right\rangle}
\ncom{\ldef}{\coloneqq}	
\ncom{\up}{\uparrow}
\ncom{\down}{\downarrow}
\ncom{\pee}{\mathds{P}}								%special P for probability functions

\DeclareMathOperator{\een}{\mathds{1}}

\begin{document}

\title{Bell Inequality Violation and Relativity of Pre- and Postselection}
\author{G. Bacciagaluppi}
\affiliation{Department of Mathematics, Utrecht University}
\author{R. Hermens}
\affiliation{Department of Mathematics, Utrecht University}
%\pacs{03.65.Ta,03.67.-a} 

\begin{abstract}
The Bell inequalities can be violated by postselecting on the results of a measurement of the Bell states. 
If information about the original state preparation is available, we point out how the violation can be reproduced classically by postselecting on the basis of this information. 
%If information about the original state preparation is available, we point out how the violation can be reproduced also by postselecting on this classical information. 
We thus propose a variant of existing experiments that rules out such alternative explanations, by having the preparation and the postselection at spacelike separation. 
Unlike the timelike case where one can sharply distinguish Bell inequality violations based on pre- or postselection of a Bell state, in our scenario the distinction between these physical effects becomes foliation-dependent. We call this `relativity of pre- and postselection'.

\end{abstract}

\maketitle

\section{Introduction} 
There are a number of remarkable effects of quantum postselection. 
Often these effects occur due to a combination with preselection as in the three-box paradox \cite{AharonovVaidman91}. 
Here we treat the case where postselection can give rise to violations of the Bell inequalities as proposed in \cite{Cohen99,Peres00} and realized experimentally in \cite{Ma12} (see also \cite{Ma16}).
This should be distinguished from the standard Bell inequality violations due to entanglement. 

In this paper we demonstrate that these violations can be reproduced by classical means, even allowing for a saturation of the superquantum bound $S=4$, by postselecting on information about the preparation procedure of the qubit pairs.
In the quantum case, access to such information is not required.
In order for this to be a genuine quantum effect, one needs to consider a setup where such information may reasonably be expected to not be available.
We accordingly propose a modification of the experiment in \cite{Ma12}.
The proposed experiment can be adapted to test simultaneously the violation of Bell inequalities due to postselection and the standard violation due to entanglement. 
It also adds a striking twist to the idea that entanglement is foliation-dependent.

%In this paper, we show how the violations based on postselection can be alternatively reproduced by postselecting on classical information about the preparation procedure of the qubit pairs, even saturating the superquantum bound $S=4$. 
%In order to have a genuine quantum effect, one needs to rule out such an alternative explanation. 
%We accordingly propose a modification of the experiment in \cite{Ma12}.
%The proposed experiment can be adapted to test simultaneously the violation of Bell inequalities due to postselection and the standard violation due to entanglement. 
%It also adds a striking twist to the idea that entanglement is foliation-dependent.

\section{Bell inequalities for Bell states} 
We begin by summarizing a few facts about Bell inequality violations.
For a pair of qubits consider the local observables
\begin{equation}
	A_i=(P_{\alpha_i}-P_{\alpha_i}^\bot)\otimes\een,~
	B_j=\een\otimes(P_{\beta_j}-P_{\beta_j}^\bot)
\end{equation}
with
\begin{equation}
	P_\varphi=\begin{pmatrix} \cos^2\varphi & \cos\varphi\sin\varphi\\ \cos\varphi\sin\varphi & \sin^2\varphi\end{pmatrix} \, .
\end{equation}	
Next, consider the four CHSH inequalities 
\begin{equation}\label{CHSH}
\begin{split}
	S_1^\psi&=\left|E^\psi_{1,1}+E^\psi_{1,2}+E^\psi_{2,1}-E^\psi_{2,2}\right|\leq2 \,,\\
	S_2^\psi&=\left|E^\psi_{1,1}+E^\psi_{1,2}-E^\psi_{2,1}+E^\psi_{2,2}\right|\leq2 \,,\\
	S_3^\psi&=\left|E^\psi_{1,1}-E^\psi_{1,2}+E^\psi_{2,1}+E^\psi_{2,2}\right|\leq2 \,,\\
	S_4^\psi&=\left|-E^\psi_{1,1}+E^\psi_{1,2}+E^\psi_{2,1}+E^\psi_{2,2}\right|\leq2 \,,
\end{split}
\end{equation}
with
\begin{equation}
	E^\psi_{i,j}=p^\psi\left(A_i=B_j\right)-p^\psi\left(A_i\neq B_j\right)\,. 
\end{equation}
Maximal violations of these inequalities can be obtained with each of the Bell states
\begin{equation}\label{Bellstates}
\begin{split}
	\ket{\Phi^+}&=\tfrac{1}{\sqrt{2}}\left(\ket{\up\up}+\ket{\down\down}\right) \,,\\
	\ket{\Psi^+}&=\tfrac{1}{\sqrt{2}}\left(\ket{\up\down}+\ket{\down\up}\right) \,,\\
	\ket{\Phi^-}&=\tfrac{1}{\sqrt{2}}\left(\ket{\up\up}-\ket{\down\down}\right) \,,\\
	\ket{\Psi^-}&=\tfrac{1}{\sqrt{2}}\left(\ket{\up\down}-\ket{\down\up}\right) \,,
\end{split}
\end{equation}
which yield the following probabilities:
\begin{equation}\label{Bellprobs} 
\begin{aligned}
	p^{\Phi^+}_{i,j}&=\cos^2(\alpha_i-\beta_j)\, ,& 
	p^{\Psi^+}_{i,j}&=\sin^2(\alpha_i+\beta_j)\, ,\\
	p^{\Phi^-}_{i,j}&=\cos^2(\alpha_i+\beta_j)\, ,&
	p^{\Psi^-}_{i,j}&=\sin^2(\alpha_i-\beta_j) \, ,
\end{aligned}
\end{equation}
where $p_{i,j}^\psi=p^\psi(A_i=B_j)$. At most one CHSH inequality can be violated for each Bell state. For sake of definiteness, we fix the spin directions 
\begin{equation}\label{directions}
	\alpha_1=0,~\alpha_2=\tfrac{\pi}{4},~\beta_1=\tfrac{\pi}{8},~\beta_2=-\tfrac{\pi}{8} \,,
\end{equation}
which give us the violations
\begin{equation}
	S_1^{\Phi^+}=S_2^{\Psi^+}=S_2^{\Phi^-}=S_1^{\Psi^-}=2\sqrt{2} \,.
\end{equation}
For the other combinations we have $S^\psi_i=0$. 

%From \eqref{Bellprobs}, one easily shows that all CHSH inequalities are satisfied by equal mixtures of any two Bell states, in particular by the perfectly correlated or anticorrelated mixtures 
%\begin{equation}\label{nonmaximal}
%\begin{split}
%	\tfrac{1}{2}P^\Phi&=\tfrac{1}{2}(\ket{\Phi^+}\bra{\Phi^+}+\ket{\Phi^-}\bra{\Phi^-}) \, , \\
%	\tfrac{1}{2}P^\Psi&=\tfrac{1}{2}(\ket{\Psi^+}\bra{\Psi^+}+\ket{\Psi^-}\bra{\Psi^-}) \, .
%\end{split}
%\end{equation}
%One also sees that $E^{\Phi^+}_{i,j}=-E^{\Psi^-}_{i,j}$ and $E^{\Phi^-}_{i,j}=-E^{\Psi^+}_{i,j}$. Thus, $\ket{\Phi^+}$ and $\ket{\Psi^-}$ always violate the same CHSH inequalities, and all $S_i$ vanish for equal mixtures of these states. The same is true for $\ket{\Phi^-}$ and $\ket{\Psi^+}$.
%Finally, an equal mixture of all four Bell states is the maximally mixed state, since the Bell states form an orthonormal basis for two qubits.
 
It follows that if Vicky prepares an equal mixture of the Bell states \eqref{Bellstates}, and sends the particles to Alice and Bob to perform measurements along the directions \eqref{directions}, each of the four subensembles leads to a maximal violation of either $S_1$ or $S_2$, but the total ensemble exhibits no correlations whatsoever.

\section{Bell Inequalities with Postselection}\label{reverse}

\subsection{(a) Quantum Case}
Let Alice and Bob independently prepare qubits.
Alice's method of preparation consists in measuring either $A_1$ or $A_2$ on a qubit in the maximally mixed state and Bob's in measuring either $B_1$ or $B_2$.
For this ensemble of qubit pairs we have of course that the outcomes of Alice's and Bob's measurements satisfy $S_i=0$ for all $i$.
Their qubits are then sent to Vicky in pairs (one from Alice and one from Bob).

Now Vicky performs on each pair a measurement of the Bell basis \eqref{Bellstates}.
Based on the outcome of this measurement Vicky constructs four subensembles of pairs of qubits.
For each of these subensembles, the outcomes of Alice's and Bob's measurement do violate one of the CHSH inequalities.
This follows simply because of the symmetry of transition probabilities; thus violation of the Bell inequalities for this case is mathematically equivalent to the standard case in which Alice and Bob perform their measurements on preselected pairs of qubits in Bell states \cite{Cohen99}. 

Although in this scenario each individual pair of qubits is in a product of eigenstates of either $A_1$ or $A_2$ and either $B_1$ or $B_2$, Vicky has no access to this information unless Alice and Bob send it classically, and it plays no role in Vicky's postselection procedure. 
As we shall see, however, classically this information could be used in principle to define an alternative postselection procedure that also leads to violation of the Bell inequalities.
%As we shall see, however, this classical information could in principle be used to define an alternative postselection procedure that also leads to violation of the Bell inequalities.

\subsection{(b) Classical Simulation}
Vicky's task is to subdivide the above totally uncorrelated ensemble into four subsensembles (with the same marginals) that violate the Bell inequalities, using information about which individual states Alice and Bob have prepared. 
For extra vividness, our initial uncorrelated ensemble will also be purely classical.
Suppose Alice chooses between either flipping a U.S. quarter dollar or a Japanese 100 yen piece, and Bob flips either a 50 euro cents or a British 10 pence \footnote{The suggested coins all have roughly similar dimensions, weight and value, but the simulation can be performed also with other currencies and denominations.}. 
They will get pairs of results with the following distributions (with `$=$' for two heads or two tails, and `$\neq$' for one head and one tail, and $a,b,c,d$ the proportions in which the four combinations of coins are flipped):
\begin{equation}\label{coindist}
	\begin{tabular}{cccccc}
 	&		& \multicolumn{4}{c}{Coins Tossed}\\
	\multirow{4}{*}{\rotatebox[origin=c]{90}{Outcomes}}
 	&	 	& \$\euro 			& \$\pounds 		& \textyen\euro 		& \textyen\pounds \\
 	%&&&&\\[-2ex]
	\cline{3-6}
	& $=$ 	& \multicolumn{1}{|c|}{$\frac{a}{2}$}        &  \multicolumn{1}{|c|}{$\tfrac{b}{2}$}       
			&  \multicolumn{1}{|c|}{$\tfrac{c}{2}$}       &  \multicolumn{1}{|c|}{$\tfrac{d}{2}$} \\
	\cline{3-6} 
 	& $\neq$ 	& \multicolumn{1}{|c|}{$\frac{a}{2}$}        &  \multicolumn{1}{|c|}{$\tfrac{b}{2}$}       
			&  \multicolumn{1}{|c|}{$\tfrac{c}{2}$}       &  \multicolumn{1}{|c|}{$\tfrac{d}{2}$} \\
	\cline{3-6} 
	&&&&&
	\end{tabular}
\end{equation}

Now let Vicky take, say, the top-left subsensemble in \eqref{coindist} ($\$=\text{\euro}$) and subdivide it at random into four subensembles in the following proportions:
\begin{equation}
	\frac{a}{4}p_{1,1}^{\Phi^+}+\frac{a}{4}p_{1,1}^{\Psi^+}+\frac{a}{4}p_{1,1}^{\Phi^-}+\frac{a}{4}p_{1,1}^{\Psi^-}=\frac{a}{2}\,,
\end{equation}
and similarly with all other boxes in \eqref{coindist}.
Note that here the $p^\psi_{i,j}$ are just theoretical numbers derived from quantum mechanics to determine the size of the subensembles.
Vicky's procedure is completely classical.
Vicky then collects the resulting pairs together in the following four subensembles:
\begin{equation}
	\begin{tabular}{ccccc}
 			 & \$\euro 			& \$\pounds 		& \textyen\euro 		& \textyen\pounds \\
 	%&&&&\\[-2ex]
	\cline{2-5}
	$=$	 	& \multicolumn{1}{|c|}{$\frac{a}{4}p^{\Phi^+}_{1,1}$}        &  \multicolumn{1}{|c|}{$\frac{b}{4}p^{\Phi^+}_{1,2}$}       
			&  \multicolumn{1}{|c|}{$\frac{c}{4}p^{\Phi^+}_{2,1}$}       &  \multicolumn{1}{|c|}{$\frac{d}{4}p^{\Phi^+}_{2,2}$} \\
	\cline{2-5} 
 	$\neq$ 	& \multicolumn{1}{|c|}{$\frac{a}{4}q^{\Phi^+}_{1,1}$}        &  \multicolumn{1}{|c|}{$\frac{b}{4}q^{\Phi^+}_{1,2}$}       
			&  \multicolumn{1}{|c|}{$\frac{c}{4}q^{\Phi^+}_{2,1}$}       &  \multicolumn{1}{|c|}{$\tfrac{d}{4}q^{\Phi^+}_{2,2}$} \\
	\cline{2-5} 
	%&&&&
	\end{tabular}\tag{I}
\end{equation} 
\begin{equation}
	\begin{tabular}{ccccc}
 			 & \$\euro 			& \$\pounds 		& \textyen\euro 		& \textyen\pounds \\
 	%&&&&\\[-2ex]
	\cline{2-5}
	$=$	 	& \multicolumn{1}{|c|}{$\frac{a}{4}p^{\Psi^+}_{1,1}$}        &  \multicolumn{1}{|c|}{$\frac{b}{4}p^{\Psi^+}_{1,2}$}       
			&  \multicolumn{1}{|c|}{$\frac{c}{4}p^{\Psi^+}_{2,1}$}       &  \multicolumn{1}{|c|}{$\frac{d}{4}p^{\Psi^+}_{2,2}$} \\
	\cline{2-5} 
 	$\neq$ 	& \multicolumn{1}{|c|}{$\frac{a}{4}q^{\Psi^+}_{1,1}$}        &  \multicolumn{1}{|c|}{$\frac{b}{4}q^{\Psi^+}_{1,2}$}       
			&  \multicolumn{1}{|c|}{$\frac{c}{4}q^{\Psi^+}_{2,1}$}       &  \multicolumn{1}{|c|}{$\tfrac{d}{4}q^{\Psi^+}_{2,2}$} \\
	\cline{2-5} 
	%&&&&
	\end{tabular}\tag{II}
\end{equation} 
\begin{equation}
	\begin{tabular}{ccccc}
 			 & \$\euro 			& \$\pounds 		& \textyen\euro 		& \textyen\pounds \\
 	%&&&&\\[-2ex]
	\cline{2-5}
	$=$	 	& \multicolumn{1}{|c|}{$\frac{a}{4}p^{\Phi^-}_{1,1}$}        &  \multicolumn{1}{|c|}{$\frac{b}{4}p^{\Phi^-}_{1,2}$}       
			&  \multicolumn{1}{|c|}{$\frac{c}{4}p^{\Phi^-}_{2,1}$}       &  \multicolumn{1}{|c|}{$\frac{d}{4}p^{\Phi^-}_{2,2}$} \\
	\cline{2-5} 
 	$\neq$ 	& \multicolumn{1}{|c|}{$\frac{a}{4}q^{\Phi^-}_{1,1}$}        &  \multicolumn{1}{|c|}{$\frac{b}{4}q^{\Phi^-}_{1,2}$}       
			&  \multicolumn{1}{|c|}{$\frac{c}{4}q^{\Phi^-}_{2,1}$}       &  \multicolumn{1}{|c|}{$\tfrac{d}{4}q^{\Phi^-}_{2,2}$} \\
	\cline{2-5} 
	%&&&&
	\end{tabular}\tag{III}
\end{equation} 
\begin{equation}
	\begin{tabular}{ccccc}
 			 & \$\euro 			& \$\pounds 		& \textyen\euro 		& \textyen\pounds \\
 	%&&&&\\[-2ex]
	\cline{2-5}
	$=$	 	& \multicolumn{1}{|c|}{$\frac{a}{4}p^{\Psi^-}_{1,1}$}        &  \multicolumn{1}{|c|}{$\frac{b}{4}p^{\Psi^-}_{1,2}$}       
			&  \multicolumn{1}{|c|}{$\frac{c}{4}p^{\Psi^-}_{2,1}$}       &  \multicolumn{1}{|c|}{$\frac{d}{4}p^{\Psi^-}_{2,2}$} \\
	\cline{2-5} 
 	$\neq$ 	& \multicolumn{1}{|c|}{$\frac{a}{4}q^{\Psi^-}_{1,1}$}        &  \multicolumn{1}{|c|}{$\frac{b}{4}q^{\Psi^-}_{1,2}$}       
			&  \multicolumn{1}{|c|}{$\frac{c}{4}q^{\Psi^-}_{2,1}$}       &  \multicolumn{1}{|c|}{$\tfrac{d}{4}q^{\Psi^-}_{2,2}$} \\
	\cline{2-5} 
	%&&&&
	\end{tabular}\tag{IV}
\end{equation} 
where $q^\psi_{i,j}=1-p^\psi_{i,j}$. 
These postselected subensembles of pairs of classical coins reproduce \emph{exactly the same} maximal violations of the Bell inequalities as in the quantum case above.

In fact, Vicky can do even better, and select instead subensembles of the form:
\begin{equation}
	\begin{tabular}{ccccc}
 			 & \$\euro 			& \$\pounds 		& \textyen\euro 		& \textyen\pounds \\
 	%&&&&\\[-2ex]
	\cline{2-5}
	$=$	 	& \multicolumn{1}{|c|}{$\tfrac{a}{4}$}        &  \multicolumn{1}{|c|}{$\tfrac{b}{4}$}       
			&  \multicolumn{1}{|c|}{$\tfrac{c}{4}$}       &  \multicolumn{1}{|c|}{$0$} \\
	\cline{2-5} 
 	$\neq$ 	& \multicolumn{1}{|c|}{$0$}        &  \multicolumn{1}{|c|}{$0$}       
			&  \multicolumn{1}{|c|}{$0$}       &  \multicolumn{1}{|c|}{$\tfrac{d}{4}$} \\
	\cline{2-5} 
	%&&&&
	\end{tabular}
		\quad
	\begin{tabular}{ccccc}
 			 & \$\euro 			& \$\pounds 		& \textyen\euro 		& \textyen\pounds \\
 	%&&&&\\[-2ex]
	\cline{2-5}
	$=$	 	& \multicolumn{1}{|c|}{$\tfrac{a}{4}$}        &  \multicolumn{1}{|c|}{$\tfrac{b}{4}$}       
			&  \multicolumn{1}{|c|}{$0$}       &  \multicolumn{1}{|c|}{$\tfrac{d}{4}$} \\
	\cline{2-5} 
 	$\neq$ 	& \multicolumn{1}{|c|}{$0$}        &  \multicolumn{1}{|c|}{$0$}       
			&  \multicolumn{1}{|c|}{$\tfrac{c}{4}$}       &  \multicolumn{1}{|c|}{$0$} \\
	\cline{2-5} 
	%&&&&
	\end{tabular}\tag{I$'$--II$'$}
\end{equation} 
\begin{equation}
	\begin{tabular}{ccccc}
 			 & \$\euro 			& \$\pounds 		& \textyen\euro 		& \textyen\pounds \\
 	%&&&&\\[-2ex]
	\cline{2-5}
	$=$	 	& \multicolumn{1}{|c|}{$0$}        &  \multicolumn{1}{|c|}{$0$}       
			&  \multicolumn{1}{|c|}{$\tfrac{c}{4}$}       &  \multicolumn{1}{|c|}{$0$} \\
	\cline{2-5} 
 	$\neq$ 	& \multicolumn{1}{|c|}{$\tfrac{a}{4}$}        &  \multicolumn{1}{|c|}{$\tfrac{b}{4}$}       
			&  \multicolumn{1}{|c|}{$0$}       &  \multicolumn{1}{|c|}{$\tfrac{d}{4}$} \\
	\cline{2-5} 
	%&&&&
	\end{tabular}
		\quad	
	\begin{tabular}{ccccc}
 			 & \$\euro 			& \$\pounds 		& \textyen\euro 		& \textyen\pounds \\
 	%&&&&\\[-2ex]
	\cline{2-5}
	$=$	 	& \multicolumn{1}{|c|}{$0$}        &  \multicolumn{1}{|c|}{$0$}       
			&  \multicolumn{1}{|c|}{$0$}       &  \multicolumn{1}{|c|}{$\tfrac{d}{4}$} \\
	\cline{2-5} 
 	$\neq$ 	& \multicolumn{1}{|c|}{$\tfrac{a}{4}$}        &  \multicolumn{1}{|c|}{$\tfrac{b}{4}$}       
			&  \multicolumn{1}{|c|}{$\tfrac{c}{4}$}       &  \multicolumn{1}{|c|}{$0$} \\
	\cline{2-5} 
	%&&&&
	\end{tabular}\tag{III$'$--IV$'$}
\end{equation} 
These subensembles now violate the same Bell inequalities with $S_i=4$. 
Thus, it is possible not only to classically simulate the quantum violations using postselection, but even to obtain superquantum violations.
%Thus, it is possible not only to simulate the quantum violations postselecting on classical information, but even superquantum violations.

\subsection{(c) Discriminating the Quantum and Classical Cases}
The structure common to the quantum and classical cases is that Vicky performs a postselection on pairs of systems in a mixture of the end products of one of two possible binary measurements.
In the classical simulation the postselection protocol makes use of the fact that Vicky has full information about the eight subensembles in \eqref{coindist}. 
For each coin pair, Vicky knows which coins were flipped and what the outcomes of these flips were.
This information is not required in the quantum case. 
There Vicky only uses the outcome of the Bell measurement to construct the subensembles.

In fact, information about which measurements Alice and Bob performed and what their outcomes were is not even available in the quantum case;
no quantum measurement on the qubit pair will help Vicky make a better guess as to what went on in Alice's and Bob's lab.
This is because from Vicky's perspective the measurements performed by Alice and Bob are fiducial and the state of the qubit pair arriving at Vicky is just the maximally mixed state.
If we place a similar restriction on the classical scenario, the possibility of violating the CHSH inequalities disappears.
The proper analogue is that instead of reporting to Vicky which coins were flipped and which outcomes were obtained, Alice and Bob just toss the coin they flipped back into the box with the coin they didn't flip and then send the box of coins to Vicky.
No measurement on the coins will reveal which one was flipped or what the outcome was \footnote{Modulo some wear and tear or DNA traces.}.

What is specifically quantum is that, although any measurement on the qubit pair will be independent of both which measurements Alice and Bob performed and which outcomes they obtained, quantum measurements are not generally independent of possible correlations between the outcomes of Alice and Bob conditional on the settings.
This is precisely what the violations of the CHSH inequalities express.

However, classical explanations of the phenomenon are not entirely ruled out.
It is conceivable that by adding hidden variables to the qubits, they do contain information about which measurements Alice and Bob performed and what their outcomes were.
What is then quantum from this perspective, is that there appears to be some limitation on possible measurements that prevents us from revealing these hidden variables.

One way to rule out such an explanation is by assuming \emph{preparation noncontextuality} \cite{Spekkens05}.
This assumption essentially boils down to the demand that because whatever Alice and Bob do, the end result for Vicky is the same (maximally mixed) quantum state, also at the hidden variable level the actions of Alice and Bob leave no discernible trace in the qubit pair.
Such an assumption may seem unreasonably strong though \cite{LeiferMaroney13} and in the next section we therefore propose an alternate experimental setup to rule out classical explanations of CHSH violations using postselection.

%Consider an experimental realization of the quantum case. Vicky performs a quantum measurement of the four Bell states \eqref{Bellstates}, and postselects four subensembles accordingly. 
%This is a striking quantum effect, but so is the violation of the Bell inequalities using classical coins. 
%And the latter provides a loophole for an alternative explanation: some local hidden variables mechanism could exploit the information about what states Alice and Bob have prepared. 
%After all, in this scenario Vicky's measurement takes place in the future light cone of Alice's and Bob's preparations and information about their settings and outcomes is in principle available at Vicky's site, indeed we can think of the qubits themselves as carrying that information. 
%A hidden variables mechanism could thus determine the results of Vicky's Bell measurement, and mimic the violation of the Bell inequailties using the procedure just described.
%[ADD A SPACETIME DIAGRAM OF THE FIRST SCENARIO?]

\section{Experimental test}\label{test}
\begin{figure} 
\includegraphics[width=8.6cm]{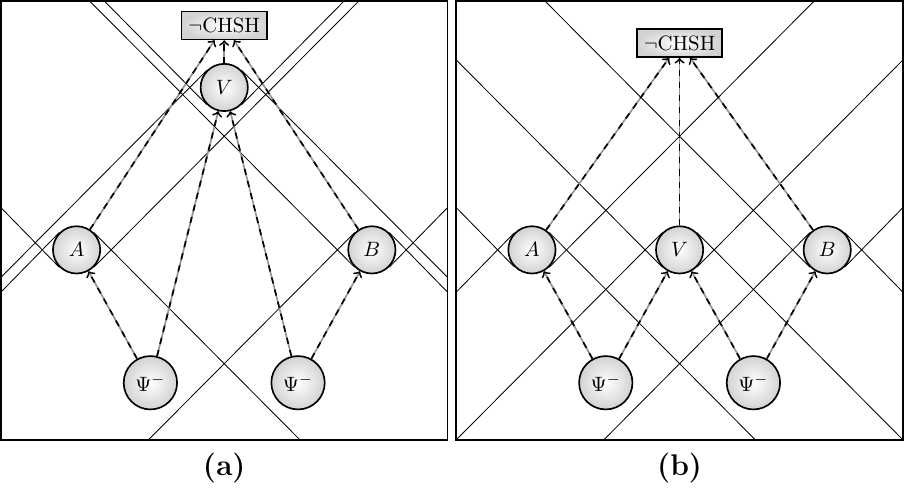}
\caption{Spacetime diagram for delayed-choice entanglement swapping (a), and for our proposed experiment (b).}\label{Figuur}
\end{figure}

%To close this loophole, we use an alternative but equivalent preparation procedure for Vicky's pairs of qubits, drawing inspiration from a protocol that was used in the experimental violation of Bell inequalities using postselection by Ma \emph{et al.}\ \cite{Ma12}.
For our proposal we draw inspiration from a protocol that was used in the experimental violation of Bell inequalities using postselection by Ma \emph{et al.}\ \cite{Ma12}.
In Ma \emph{et al.}'s protocol, each of the qubits in Vicky's uncorrelated pair is part of a pair prepared in the state $\ket{\Psi^-}$.
Vicky shares one of these pairs with Alice, and the other with Bob (see \autoref{Figuur} (a)). Alice and Bob then perform their local experiments on their qubits. 
If we think of Alice's and Bob's measurements as collapsing the state also at Vicky's site, this procedure leads to the same mixed states for Vicky's qubits as in the previous scenario. 
The difference is that Alice and Bob have now prepared them at a distance.

Initially, of course, Alice's and Bob's results will be completely uncorrelated.
At an arbitrary point in the future, however, Vicky can decide to perform a measurement in the Bell basis \eqref{Bellstates}.
This is delayed-choice entanglement swapping \cite{Peres00}.
The outcomes of this measurement can then be used to postselect subensembles for which the measurement results of Alice and Bob become correlated and violate a CHSH inequality.

When Vicky's measurement is timelike separated from both Alice's and Bob's measurements as in \autoref{Figuur} (a), the explanation of the CHSH violation is unambiguously due to postselection.
Although the experiment by Ma \emph{et al.}\ \cite{Ma12} was accordingly set up to ensure timelike separation between Alice and Bob's and Vicky's measurements, it is precisely this feature that provides the loophole for a classical explanation of the results.
In order to ensure that information about Alice's and Bob's measurements %preparations 
cannot reach Vicky's site, and thus that the experimental violation of the Bell inequalities due to postselection is a genuine quantum effect, it is not necessary that Alice and Bob are at spacelike separation from each other, but we need to make sure that their measurements (including their choice of settings) are at spacelike separation from Vicky's.  
We thus propose this as a modification of the Ma \emph{et al.}\ experiment, as in \autoref{Figuur} (b).

\section{Relativity of pre- and postselection}\label{relative}
In quantum theory, spacelike separated measurements commute. This is the basis for what Shimony has called the `peaceful coexistence' of quantum theory and relativity \cite{Shimony78,Shimony86}.
But there is of course a tension with the idea that quantum state collapse occurs instantaneously across space.
A proposal to resolve this tension is to embrace the idea that quantum states are defined on spacelike hypersurfaces and encode the probabilities for results of measurements to the future of the given hypersurface conditional on results of measurements to its past \cite{Fleming86,Fleming89,Fleming96,Myrvold00,Myrvold02}.
Consequently, entanglement of distant particles becomes a foliation-dependent notion: while the probabilities for Alice's and Bob's results are invariant, whether a qubit pair is entangled when Alice performs a measurement depends on the time order between their measurements.
To capture this phenomenon, Myrvold \cite{Myrvold00,Myrvold02} has coined the term `relativity of entanglement'. 

\begin{figure}[h] 
\includegraphics[width=8.6cm]{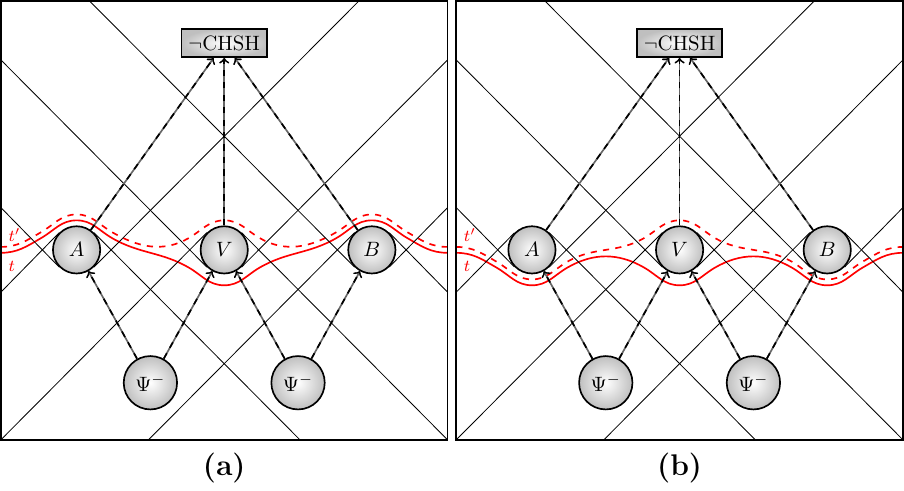}
\caption{Spacetime diagrams with foliations with different time ordering of events. In (a) Vicky's measurement postselects sub-ensembles while in (b) the measurement acts as a preslection procedure.}\label{RelPrePost}
\end{figure}

The experiment we propose adds a further dramatic touch to this idea. Because of the spacelike separation between Vicky's measurements and Alice and Bob's, the same experiment can be alternatively described in two different ways: either as Vicky performing a series of Bell measurements on maximally mixed pairs prepared by Alice and Bob (\autoref{RelPrePost} (a)), or as Alice and Bob performing a series of EPR measurements on maximally entangled pairs prepared by Vicky (\autoref{RelPrePost} (b)). 
In other words, depending on the choice of foliation, Vicky's measurement acts as a preselection or a postselection. 
We now have \emph{relativity of pre- and postselection} \footnote{Since any Bell measurement consists of entangling a pair of particles with an ancilla and performing a measurement on the latter, already Cohen  \cite{Cohen99} points out that the difference between pre- and postselection can be reduced to the timing of the measurement on the ancilla. But he cashes this out differently: postselection for him reflects the `counterfactual entanglement' that would have existed had we preselected instead.}. 

This addresses one potential objection to our analysis in the previous section. 
If we imagine that Alice's and Bob's measurements actually collapse the state at a distance also at Vicky's site, then the individual pairs of qubits on which Vicky performs the Bell measurements are in definite product states. 
Thus, although there are no quantum-mechanical measurements Vicky can perform that will reveal the individual states of the qubits, the qubits themselves carry information that is perfectly correlated with the information about Alice's and Bob's measurements, and a hidden variables mechanism might exploit it. %RH: removed "classical"
However, this mechanism requires a preferred foliation in which Alice's and Bob's measurements take place before Vicky's.

\begin{figure}[hb] 
\includegraphics[width=8.6cm]{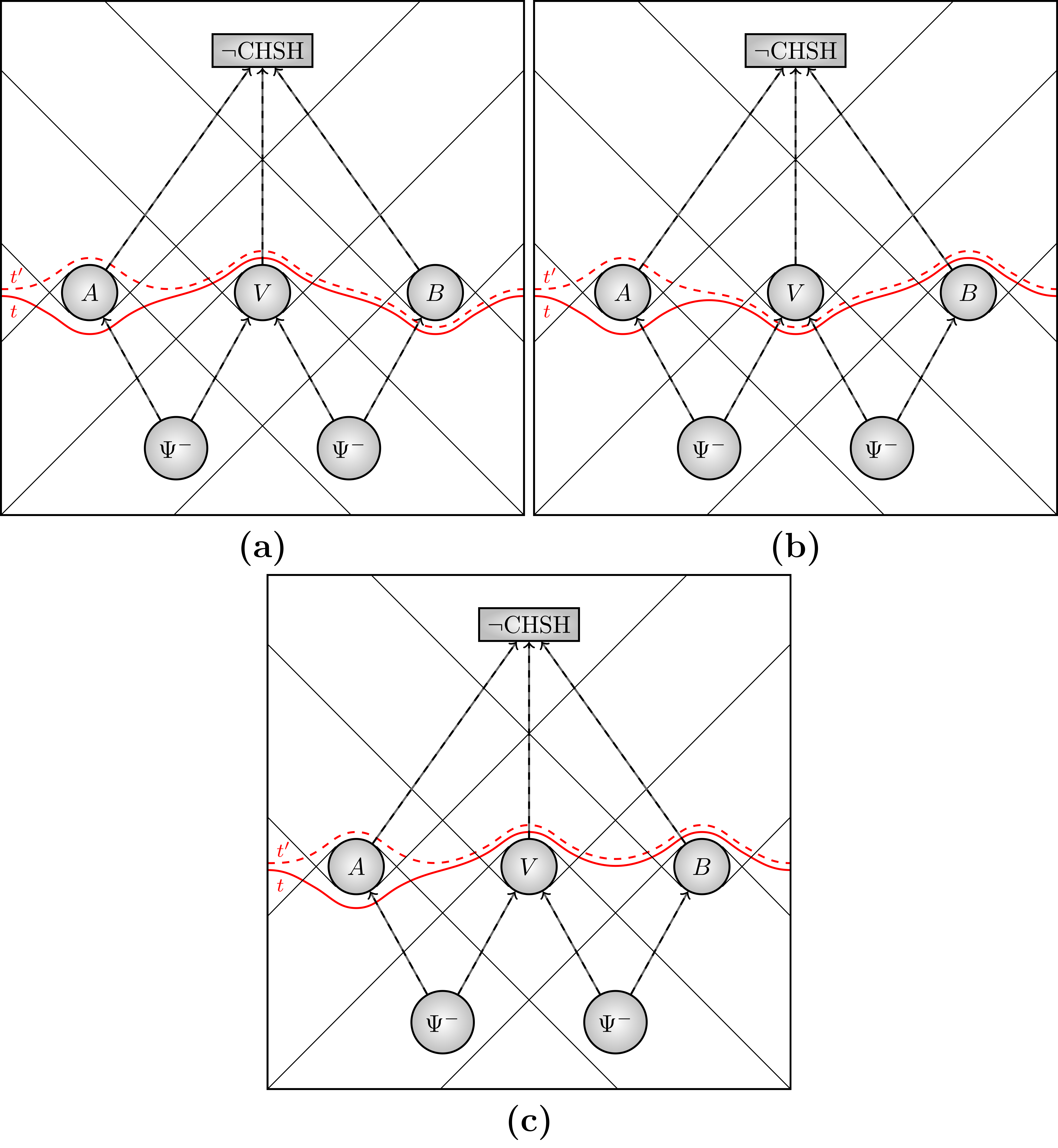}
\caption{Spacetime diagrams with foliations with different time ordering of events. In (a) Alice performs a measurement on a qubit entangled with the qubit at B, in (b) it is entangled with the qubit in V, and in (c) it is not entangled.}\label{RelEnt}
\end{figure}

Finally, the relativity of pre-and postselection suggests considering the case in which all three measurements are at spacelike separation from each other as indeed shown in \autoref{Figuur} (b)\footnote{Note that in the Ma \emph{et al.}\ experiment, Alice and Bob are already spacelike separated. The setup with all three parties at spacelike separation has previously been proposed in \citep{Cabello12} as a modified test of nonlocality, but not specifically in the context of postselection.}. 
By choosing an appropriate foliation, the same three measurements can be given any arbitrary time order. 
Thus in this scenario, not only does the choice of foliation affect whether Alice performs a measurement on an entangled qubit or not, it also affects with which other qubit it is entangled (\autoref{RelEnt}) \footnote{It is also possible to choose the foliation such that the creation of Alice's qubit occurs after Bob's measurement. The scenario in which these events are instead timelike separated was realized experimentally in \cite{Megidish13}.}.
The experiment can now be seen both as a modification of the delayed-choice entanglement swapping by Ma \emph{et al.}, and as a modification
of the loophole-free Bell-EPR experiment by Hensen \emph{et al.}\ \cite{Hensen15}, where Vicky's measurement is part of the preparation procedure of Alice's and Bob's qubits \footnote{Note that in the Hensen \emph{et al.}\  experiment, the entanglement-swapping Bell measurement takes place at spacelike separation from the choice of settings at Alice's and Bob's locations, so as to rule out a causal mechanism inducing settings-source dependence. But Alice's and Bob's measurements, although begun at spacelike separation from the Bell measurement, are completed in its future light cone (\cite{Hensen15}, Figure 2). Thus, there are spacetime foliations  % -- or frames of reference, given the geometry of the experiment (\cite{Hensen15}, Figure 1) --
in which the EPR measurements are performed from beginning to end in the future of the Bell measurement, but no foliations in which the Bell measurement is performed in the future of the entire EPR measurements, and the violation of the Bell inequalities is unambiguously due to entanglement.}. 
In this version, the experiment becomes a (loophole-free) \emph{simultaneous test} of Bell inequality violations due to entanglement and to postselection.

\section{Conclusion}
The quantum-mechanical predictions are invariant under change of foliation, because measurements at spacelike separation commute. Because of the relativity of pre- and postselection, instead, the difference between Bell inequality violations due to entanglement and due to postselection is no longer invariant. What in the case of timelike separation appear as physically different effects, in the case of spacelike separation turn out to be one and the same physical effect. 

When in 1905 Einstein related two seemingly very different effects in the introduction to his `On the electrodynamics of moving bodies' \cite{Einstein05}, it led to the unification of electric and magnetic fields as one single physical object. 
Perhaps the relativity of pre-and postselection in violations of the Bell inequalities is trying to tell us that the very notion of quantum state is in need of equally deep revision. 
This is indeed what Abner Shimony (1928--2015), to whose memory we wish to dedicate this paper, thought about the relativity of entanglement. 
As he eloquently put it \cite{Shimony86}: 
\begin{quote}\small
[T]he two accounts of \emph{processes} from initial to final sets of events are in disaccord. But it is important to note
that the process is a theoretical construction. [...] The thesis of peaceful coexistence presupposes a conceptually
coherent reconciliation of the descriptions from the standpoints of [the frames] $\Sigma$ and $\Sigma'$. Even more desirable, in the spirit of
the geometrical formulation of space-time theory, would
be a coordinate-free account.
\end{quote}
%This paper is dedicated to the memory of Abner Shimony (1928--2015).

\section*{Acknowledgements} 
We are grateful to our Masters student Fransje Prins for getting us to think very hard about relativity of entanglement in our foundations of quantum mechanics course, as well as to Maurits Moeys for discussions of entanglement swapping in the Hensen \emph{et al.}\ experiment, and to Ad\'{a}n Cabello for discussions of the Ma \emph{et al.}\ experiment and comments on an earlier draft.
We are grateful for useful referee comments that helped us improve the paper. % and to Lev Vaidman for helpful comments on the penultimate version. %If we do send again to Phys. Rev., we should probably add Lev's name only at proof stage.
GB adds: Thanks to the ever-amazing Abner Shimony, who first introduced me to the foundations of quantum mechanics and encouraged me to pursue them. 
RH received financial support from the NWO (Veni-project nr. 275-20-070).

\bibliography{refs}

\end{document}